\documentclass[11pt,preprint]{elsarticle}
\usepackage{bm}
\newcommand{\ve}[1]{\bm{#1}}

\newcommand{\bff}{\ve{f}}
\newcommand{\bx}{\ve{x}}

\newcommand{\Rdot}{\dot{R}}
\newcommand{\Rddot}{\ddot{R}}

\newcommand{\bmu}{\vec{\boldsymbol{\mu}}}
\newcommand{\btheta}{\vec{\boldsymbol{\theta}}}

\newcommand{\bSigma}{\boldsymbol{\Sigma}}

\newcommand{\eps}{\varepsilon}

\newcommand{\vbx}{\vec{\ve{x}}}
\newcommand{\vbg}{\vec{\ve{g}}}

\newcommand{\vbf}{\vec{\ve{f}}}

\newcommand{\mom}{\mu'}
\newcommand{\bmom}{\vec{\boldsymbol{\mu}}'}

\newcommand{\dd}{\text{d}}

\newcommand\Rey{\mbox{\textit{Re}}}  
\newcommand\Ca{\mbox{\textit{Ca}}} 

\usepackage{epsf}
\usepackage{amsmath}
\usepackage{amssymb}
\usepackage{setspace}
\usepackage{enumerate}
\usepackage{graphicx}
\usepackage{float}
\usepackage{footnote}
\usepackage{scalefnt}
\usepackage{microtype}
\usepackage{xfrac}
\usepackage{transparent}
\usepackage{rotate}
\usepackage{array}
\usepackage{tabu}
\usepackage[normalsize]{subfigure}
\usepackage[mediumspace,mediumqspace,squaren]{SIunits}
\usepackage{xcolor}
\usepackage[margin=1in]{geometry}

\let\today\relax
\date{\today}

\begin{document}

\begin{frontmatter}

\author[add1]{Spencer H. Bryngelson}\corref{cor}
\cortext[cor]{Corresponding author; spencer@caltech.edu}
\author[add2]{Alexis Charalampopoulos}
\author[add2]{Themistoklis P. Sapsis}
\author[add1]{Tim Colonius}

\address[add1]{Division of Engineering and Applied Science, California Institute of Technology, Pasadena, CA 91125, USA}
\address[add2]{Department of Mechanical Engineering, Massachusetts Institute of Technology, Cambridge, MA 02139, USA}

\title{
A Gaussian moment method and its augmentation via LSTM recurrent neural networks 
for the statistics of cavitating bubble populations 
}

\begin{abstract}
    Phase-averaged dilute bubbly flow models require high-order statistical moments of the bubble population. The method of classes, which directly evolve bins of bubbles in the probability space, are accurate but computationally expensive. Moment-based methods based upon a Gaussian closure present an opportunity to accelerate this approach, particularly when the bubble size distributions are broad (polydisperse).  
    For linear bubble dynamics a Gaussian closure is exact, but for bubbles undergoing large and nonlinear oscillations, it results in a large error from misrepresented higher-order moments.
    Long short-term memory recurrent neural networks, trained on Monte Carlo truth data, are proposed to improve these model predictions. 
    The networks are used to correct the low-order moment evolution equations and improve prediction of higher-order moments based upon the low-order ones.
    Results show that the networks can reduce model errors to less than $1\%$ of their unaugmented values.
\end{abstract}


\journal{International Journal of Multiphase Flow}

\end{frontmatter}

\section{Introduction}\label{s:intro}

The dynamics of bubble clouds play a central role in diverse applications from analyzing injury from blast trauma~\citep{laksari15}, understanding kidney stone pulverization in shock- and ultrasound-based lithotripsy~\citep{pishchalnikov03,ikeda06,maeda19}, designing artificial heart valves and pumps~\citep{brennen15}, and minimizing cavitation erosion over propellers and hydrofoils~\citep{chang08}.
When the size distributions of bubble nuclei are broad, the average response of the bubbles to pressure fluctuations damps and disperses~\citep{smereka02,shimada00,colonius08,ando11}. 
Ensemble-averaged bubbly flow models~\citep{zhang94} must account for such size distributions and disequilibria if they are to represent the dynamics of realistic bubbly flows.
Current methods for representing such polydispersity are computationally expensive, even in the dilute limit~\citep{bryngelson19}.

Previous models have approximated statistical moments of these populations using the method of classes~\citep{vanni00,bryngelson19,ando11}.
This approach evolves bins of the bubble size distribution.
While straightforward, this approach is costly in a simulation environment with spatial inhomogeneities, since it involves solving a large system of equations at each point in space.
An alternative approach is Monte Carlo methods; they solve the governing equations by discretely sampling the bubble population~\citep{zhao07}.
Unfortunately using Monte Carlo for this purpose is also expensive, and thus are usually only used for validation of other methods~\citep{zucca07}.

In the present work, we explore moment methods as an alternative to the aforementioned approaches.
Moment methods evolve some parameters of a distribution, such as moments~\citep{hulburt64} or expected values~\citep{moyal49}, that follow from a population balance equation.
This technique has been used to model polydisperse bubbly flows, including coalescence and breakup~\citep{heylmun19} and dilute gas-particle flows~\citep{kong19,capecelatro13,desjardins08}, though to our knowledge has not been applied to cavitating bubble populations, which undergo large volume changes.
For nonlinear bubble dynamics the moment evolution equations cannot be expressed in terms of only lower order moments.
One way to treat this issue is the quadrature-based moment method (QBMM), which approximates unclosed terms by evolving quadrature points and weights that correspond to an assumed-underlying distribution (often Gaussian)~\citep{mcgraw97,marchisio05}. 
However, QBMMs have their own difficulties, such as delta-shocks and negative quadrature node weights, that can pollute solutions~\citep{chalons12,fox09}.
Further, high-order moment predictions are still computationally expensive in this framework.

Instead, the current model evolves a multivariate probability density function that describes the bubble distribution.
Probability calculus and a Gaussian closure ansatz determine the moment evolution equations.
This approach is more general than the classes method because it utilizes random variables in the full bubble state configuration (instantaneous and equilibrium radii and radial velocity), rather than just the equilibrium bubble size parameter.
This provides additional model flexibility that can describe, for example, experimental conditions that only have statistical estimates of bubble population state.


However, this approach is potentially limited for bubble populations with significant high-order statistics. 
A more general density function can address this, though this is computationally expensive and challenging because of the so-called moment problem~\citep{akhiezer65,stieltjes94}.
Instead, recent developments suggest that a recurrent neural network (RNN) can efficiently augment such imperfect dynamical systems, accounting for the dependency of current-time data on previous data~\citep{wan18plos, wan19, srinivasan19}.
In particular, long short-term memory (LSTM) RNNs are well suited for this task as they truncate gradient-based errors when they do not affect the prediction, resulting in short training times~\citep{hochreiter1997long}
Here, we use LSTM RNNs to improve model predictions for the the low-order moment evolution and high-order moment evaluation when high-order statistics are significant.



Section~\ref{s:model} presents an overview of the bubble dynamic model and governing equations.
Section~\ref{s:pdf} formulates the density-function-assumed statistical evolution model (appendix~\ref{a:moments} includes references for specific derivations).
Section~\ref{s:linear} shows results for linear bubble dynamics of both $R_o$ monodisperse and polydisperse populations.
Section~\ref{s:nonlinear} extends this analysis to nonlinear dynamics via the Rayleigh--Plesset equation.
Section~\ref{s:ML} presents the RNN that improves model predictions and results from it.
Section~\ref{s:limit} discusses the limitations of this method and potential treatments for them.
Section~\ref{s:conclusion} concludes the paper.

\section{Bubble dynamics model}\label{s:model}

As a representative scenario, the bubbles are non-interacting, isothermal, and surface tension is neglected.
While such assumptions are not appropriate under all circumstances, this model includes the key driving dynamics and can be appended to represent additional physical effects.
The Rayleigh--Plesset equation thus represents the single-bubble dynamics:
\begin{gather}
	R \Rddot  + \frac{3}{2} \Rdot^2 + \frac{4}{\Rey} \frac{\Rdot}{R} = 
	\Ca \left[ \left( \frac{R_o}{R} \right)^{3 \gamma}  - 1 \right] - C_p 
	\label{e:RPE}
\end{gather}
where $\gamma$ is the polytropic index, $R_o$ is the equilibrium bubble radius, $R$ is the instantaneous radius, and the dots represent time derivatives.
The Reynolds number, cavitation number, and dimensionless pressure forcing are
\begin{gather}
	\Rey \equiv \sqrt{ \frac{ p_o}{\rho_o}  } \frac{R_o}{\nu_o}, \quad
	\Ca \equiv \frac{p_o - p_v}{p_o}, \quad \text{and} \quad
	C_p \equiv \frac{p_\infty - p_o}{p_o},
\end{gather}
respectively, where $\nu_o$ is the reference kinematic viscosity, $\rho_o$ is the reference liquid pressure,
and $p_v$, $p_o$, and $p_\infty$ are the vapor, ambient, and liquid far-field pressures.
The bubbles are gas-filled ($p_v = 0$ and so $\Ca = 1$) and compress adiabatically ($\gamma = 1.4$).
The time-independent pressure ratio $p_o / p_\infty$ modifies the bubble collapse strength.
In general, phase-averaged model flows have a time-dependent $p_\infty$, though the time scale of the bubble dynamics is much shorter than that of the flow that advects them in these cases. 
Thus, the time-independent case serves as a model problem, though the method presented here can extend to time-dependent pressures.

\section{Density-shape-assumed model formulation}\label{s:pdf}

\subsection{General, polydisperse model}\label{s:poly}

The polydisperse bubble dynamics of section~\ref{s:model} entail three uncertain variables: $R$, $\Rdot$, and $R_o$. 
The probability of any such state $\vbx = \{ R,\Rdot,R_o \}$ occurring is
\begin{gather}
	P = P(\vbx,\btheta,t),
\end{gather}
where $P$ is at most a trivariate probability density function with parameters (e.g.\ means, shape parameters) $\btheta$ and raw moments $\bmom$. 
There are
\begin{gather}
	\sum_{q = 1}^{N_q} \binom{N_r + q - 1}{q}
\end{gather}
such moments, where $N_r = 3$ is the number of random variables, $q$ is the moment order index, and $N_q$ is the highest moment order. 
The specific moments are 
\begin{gather}
	\mom_{lmn} = \int P R^l \Rdot^m R_o^n \, \dd \bx
    \label{e:mom}
\end{gather}
where $l + m + n = q$.

A governing equation for $P$ follows from the usual master equation  
\begin{gather}
	\frac{\dd P}{\dd t} = \frac{ \partial P}{\partial t} + 
	\frac{\partial}{\partial R} (P \Rdot ) + 
	\frac{\partial}{\partial \Rdot} (P \ddot{R} ) = 0,
    \label{e:master}
\end{gather}
where $\dot{R_o} = 0$ since the $R_o$ distribution is static~\citep{ando10}. 
This constraint complements~\eqref{e:master} as a marginal condition for $P(R,\dot{R})$ when $P(R_o)$ is specified.
The moment system evolves as
\begin{gather}
	\frac{\partial \bmom}{\partial t} = \vbf ( \bmom, \bx ),
	\label{e:system}
\end{gather}
where $\vbf$ is over all moments $\{ l,m,n \}$:
\begin{gather}
	f_{lmn} = l \mom_{l-1,m+1,n} + n \int \Rddot(\vbx) 
	R^l \Rdot^{m-1} 
	R_o^n P(\vbx,\btheta) \, \dd \vbx,
	\label{e:rhs}
\end{gather}
where $\Rddot$ follows from~\eqref{e:RPE} and the integration is over the support of $P$.
We call this the PDF-based model (or PDF) throughout.
The derivation of~\eqref{e:rhs} is in appendix~A. 
Thus, \eqref{e:system} is a nonlinear system of integro-differential equations that requires specification of $N_q$, $P$, and the transformation $\bmom \Leftrightarrow \btheta$.
The second-order accurate Adams--Bashforth method evaluates the time derivative.

\subsection{Monodispersity in $R_o$}\label{s:mono}

$R_o$-monodisperse cases test the model performance throughout.
$P(R_o) \to \delta(R_o^\ast - 1)$ describes these cases, where $\delta$ is the Dirac delta function, though these populations can still be in bubble size and velocity disequilibrium.
These cases require no $R_o$ moments ($n = 0$, $P = P(R,\Rdot,t)$, $N_r = 2$).
Thus, we compute the $R_o$ integral of~\eqref{e:rhs} analytically, resulting in double integrals over $R$ and $\Rdot$.

\section{Prediction of linear bubble dynamics}\label{s:linear}

This section considers linear bubble dynamics as a case for which a Gaussian closure is exact.
Linearizing~\eqref{e:RPE} about $R = R_o$ yields
\begin{gather}
	\Rddot + \beta(R_o) \Rdot + \omega^2(R_o) (R - R_o) = - \frac{C_p}{R_o}
\end{gather}
where
\begin{gather}
	\beta = \frac{4}{ \Rey R_o^2} 
	\quad \text{and} \quad 
	\omega^2 = \frac{ 3 \gamma \Ca }{R_o^2}
\end{gather}
characterize the damping rate and bubble natural frequency, respectively.

\subsection{$R_o$-monodisperse populations}\label{s:linear_mono}

The integrals of~\eqref{e:rhs} are evaluated analytically for linear $R_o$-monodisperse bubble populations:
\begin{gather}
\renewcommand{\arraystretch}{1.25}
	\frac{ \partial \bmom }{\partial t} = 
	\vbf = \begin{bmatrix}
		f_{100} \\ f_{010} \\ f_{200} \\ f_{020} \\ f_{110}
	\end{bmatrix}
	=
	\begin{bmatrix}
		 \mom_{010} \\ 
		 - \beta \mom_{010} - \omega^2 ( \mom_{100} + R_o ) - C_p / R_o  \\ 
		 2 \mom_{110} \\
		 -2 (\beta \mom_{020} + \omega^2 ( \mom_{110}+ R_o \mom_{010} )) \\
		 \mom_{020} - (\beta \mom_{110} + \omega^2 ( \mom_{200} + R_o \mom_{100})) 
	\end{bmatrix}
	.
\end{gather}
Since $\vbf$ requires only a finite number of moments (only the low-order moments, $\bmom$ up to $N_q = 2$), this system is closed for any five-parameter bivariate distribution $P$.
The multivariate normal distribution
\begin{gather}
    P(\vbx,\btheta) = 
        \frac{1}{2 \pi \sqrt{ | \bSigma | }} 
        \exp \left( -\frac{1}{2} (\vbx - \bmu)^\top \bSigma^{-1} (\vbx - \bmu) \right)
    \label{e:pdf}
\end{gather}
demonstrates this method, where $\bSigma$ is the covariance matrix.
For $R_o$-monodisperse cases $\vbx = \{ R, \Rdot \}$ are the random variables, $\bmu = \{\mu_R, \mu_{\dot{R}} \}$ are their means, and the transformation $\bmom \Leftrightarrow \btheta$ is
\begin{gather}
\renewcommand{\arraystretch}{1.75}
	\btheta = 
    \begin{bmatrix} 
        \mu_R \\ \sigma_R^2 \\ \mu_{\Rdot} \\ \sigma_{\Rdot}^2 \\ \rho_{R \Rdot}\vphantom{\frac{\mom_1}{\sqrt{\mom^2_{1}}}} 
    \end{bmatrix} = 
	\begin{bmatrix}  
        \mom_{100}  \\ 
	 	\mom_{200} - \mom^2_{100} \\
		\mom_{010} \\
		\mom_{020} - \mom^2_{010} \\
		\frac{ \mom_{110} - \mom_{100} \mom_{010} }  { \sqrt{ \mom_{200} - \mom^2_{100} } \sqrt{ \mom_{020} - \mom^2_{010} }}
	\end{bmatrix}	
    \quad \text{and} \quad 
	\bmom = 
    \begin{bmatrix} 
	    \mom_{100} \\ \mom_{010} \\ \mom_{200} \\ \mom_{020} \\ \mom_{110}
	\end{bmatrix} = 
    \begin{bmatrix}  
		\mu_R  \\ 
	 	\mu_{\Rdot} \\
		\mu_R^2 + \sigma_R^2 \\
		\mu_{\Rdot}^2 + \sigma_{\Rdot}^2 \\
		\mu_{R} \mu_{\Rdot} + \rho_{R \Rdot} \sigma_R \sigma_{\Rdot} 
	\end{bmatrix}	
    .
    \label{e:trans}
\end{gather}

An example linear dynamics case determines if this method can reproduce the expected statistics for a binormal distribution function.
It is specified by $R_o = \mu_R (t=0)$, $\Rey = 20$, and $p_o / p_\infty = 0.9$, with initial conditions $\mu_R = 1$, $\mu_{\Rdot} = 0$, $\sigma_R^2 = 0.01^2 R_o $, $\sigma_{\Rdot}^2 = 0.01$, and $\rho_{R \Rdot} = 0$, though the conclusions are insensitive to these choices.
The relative metric
\begin{gather}
	\eps(\ast) \equiv \frac{\lVert \ast_\mathrm{MC} - \ast_\mathrm{PDF}  \rVert_2}{ \lVert \ast_\mathrm{MC} \rVert_\infty}
\end{gather}
quantifies the error.
Subscripts MC and PDF refer to the Monte Carlo and PDF-based models, respectively, and $\lVert \ast \rVert_s$ is the $L_s$ norm.

\begin{figure}
    \centering
    \includegraphics[scale=1]{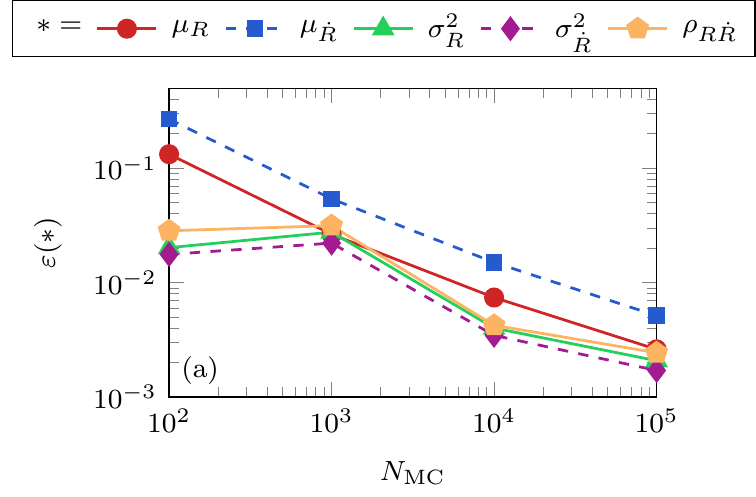} \hspace{0.2cm}
    \includegraphics[scale=1]{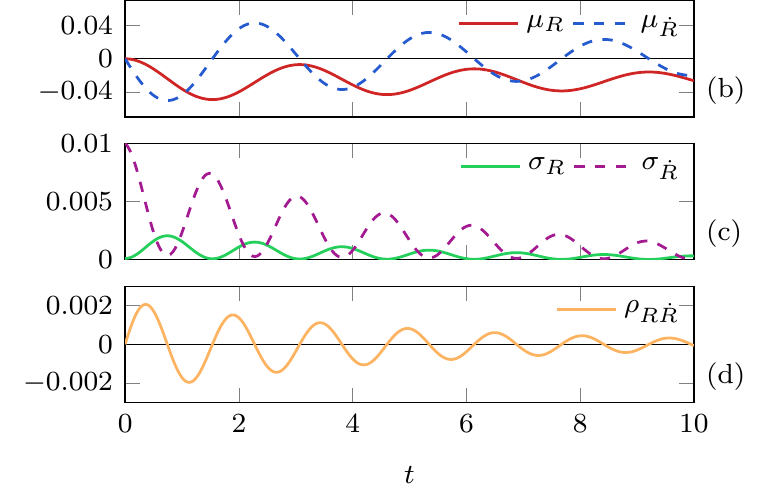}
    \caption{Linear bubble dynamics for an example case.
        (a) Validation error $\eps$ (see text) of the density function parameters $\btheta$ for varying Monte Carlo sample number $N_\text{MC}$ and (b)--(d) their temporal evolution.
        (b) shows $\mu_R - \mu_R(t=0)$ instead of $\mu_R$ to ease visualization.
    }
    \label{f:linear}
\end{figure}

Figure~\ref{f:linear}~(a) shows the low-order moment errors $\eps$ over three periods of the mean bubble dynamics.
They are small and decay with increasing $N_\mathrm{MC}$, consistent with the expected Monte Carlo sampling error for all moments, validating the linear model above. 
Figure~\ref{f:linear}~(b) shows the low-order moment evolution.
The means $\mu_\ast$ have the same dynamics as a damped harmonic oscillator and the variances $\sigma_\ast$ grow and decay out of phase.
A covariance $\rho_{R\Rdot}$ develops despite the linear dynamics and initially independently distributed random variables $\rho_{R\Rdot}(t=0) = 0$.
Thus, representing the linear bubble population statistics requires a random variable covariance parameter.


\subsection{$R_o$-polydisperse populations}\label{s:linear_poly}

A trivariate correlated normal distribution (following~\eqref{e:pdf}) is used to predict linear, polydisperse bubble dynamics.
Thus in~\eqref{e:pdf}, $\vbx = \{ R, \Rdot, R_o \}$, $\bmu = \{\mu_R, \mu_{\dot{R}}, \mu_{R_o} \}$, and $\bmom \Leftrightarrow \btheta$ follows from~\eqref{e:trans} with the additional rows:
\begin{gather}
\renewcommand{\arraystretch}{1.75}
	\btheta = 
    \begin{bmatrix} 
        \mu_{R_o} \\ \sigma_{R_o}^2 \\ 
	    \rho_{R R_o}   \vphantom{\frac{\mom_1}{\sqrt{\mom^2_{1}}}} \\ 
        \rho_{\dot{R} R_o}   \vphantom{\frac{\mom_1}{\sqrt{\mom^2_{1}}}}
    \end{bmatrix} = 
	\begin{bmatrix}  
		\mom_{001} \\
		\mom_{002} - \mom^2_{001} \\
		\frac{ \mom_{101} - \mom_{100} \mom_{001} } { \sqrt{ \mom_{200} - \mom^2_{100} }  \sqrt{ \mom_{002} - \mom^2_{001} } }  \\
		\frac{ \mom_{011} - \mom_{010} \mom_{001} } { \sqrt{ \mom_{020} - \mom^2_{010} }  \sqrt{ \mom_{002} - \mom^2_{001} } } 
	\end{bmatrix}	
        \quad \text{and} \quad 
	\bmom = 
    \begin{bmatrix} 
	    \mom_{001} \\ 
	    \mom_{002} \\
	    \mom_{101} \\ 
        \mom_{011}
	\end{bmatrix} = 
    \begin{bmatrix}  
		\mu_{R_o} \\
		\mu_{R_o}^2 + \sigma_{R_o}^2 \\
		\mu_{R} \mu_{R_o} + \rho_{R R_o} \sigma_R \sigma_{R_o} \\
		\mu_{\Rdot} \mu_{R_o} + \rho_{\Rdot R_o} \sigma_{\Rdot} \sigma_{R_o} 
	\end{bmatrix}	
    .
\end{gather}
Evaluating the integrals of~\eqref{e:rhs} via adaptive Gaussian quadrature to within $10^{-7}\%$ relative error ensures that the observed errors result from the statistical model.
The system has the same parameterization as that of section~\ref{s:linear_mono}, with additional initial conditions $\mu_{R_o} = \mu_{R}$, $\rho_{\dot{R} R_o} = 0$, and $\rho_{R R_o} = 0$.

\begin{figure}[H]
	\centering
	\includegraphics[scale=1]{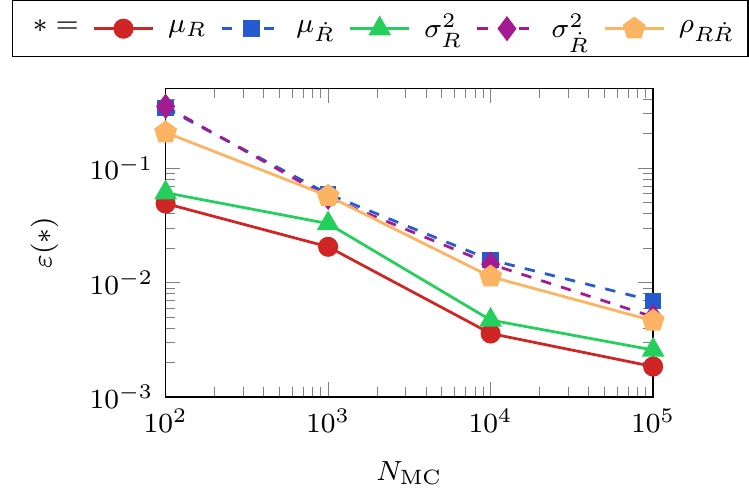} 
	\caption{Validation errors for linear, $R_o$-polydisperse bubble population dynamics.}
	\label{f:linear_poly}
\end{figure}

Figure~\ref{f:linear_poly} shows the validation error associated with a general polydisperse bubble population when comparing to a Monte Carlo simulation of varying sample sizes $N_\mathrm{MC}$.
We again compute the errors over three cycles of the mean bubble dynamics.
Similar to figure~\ref{f:linear}~(a), increasing $N_\mathrm{MC}$ results in the expected decrease in Monte Carlo sample error for all density function parameters.
These errors are of similar size to those of the $R_o$-monodisperse results for the largest $N_\mathrm{MC}$ considered.
Thus, this model can represent linear polydisperse bubble dynamics at least up to this accuracy.

\section{Prediction of nonlinear bubble dynamics}\label{s:nonlinear}

Gaussian closure is not exact for nonlinear bubble dynamics.  
This section characterizes the errors incurred by applying this closure.  
The bubbles evolve according to the Rayleigh--Plesset equation~\eqref{e:RPE} with different pressure ratios $p_o/p_\infty$.  
For small pressure ratios, the bubbles collapse violently and oscillate nonlinearly, whereas as the pressure ratio approaches unit, linear dynamics are recovered. 
Thus, the moments of such a bubble population match those of the linear case when $p_o/p_\infty \to 1$ and $\sigma_{\Rdot} \to 0$.

The initial bubble populations are independently distributed and Gaussian with means $\mu_R = 1$ and $\mu_{\Rdot} = 0$ and variances $\sigma_R^2 = 0.01$ and $\sigma_{\Rdot}^2 = 0.05$.
Monte Carlo simulations with $N_\text{MC} = 10^5$ samples serve as a surrogate for the exact solution throughout.
We restrict our analysis to $R_o$-monodisperse cases. However, as discussed in section~\ref{s:pdf} and shown in section~\ref{s:linear}, including $R_o$ polydispersity is straightforward.

\subsection{Low-order moment prediction}\label{s:loworder}

\begin{figure}
    \centering
    \includegraphics[scale=1]{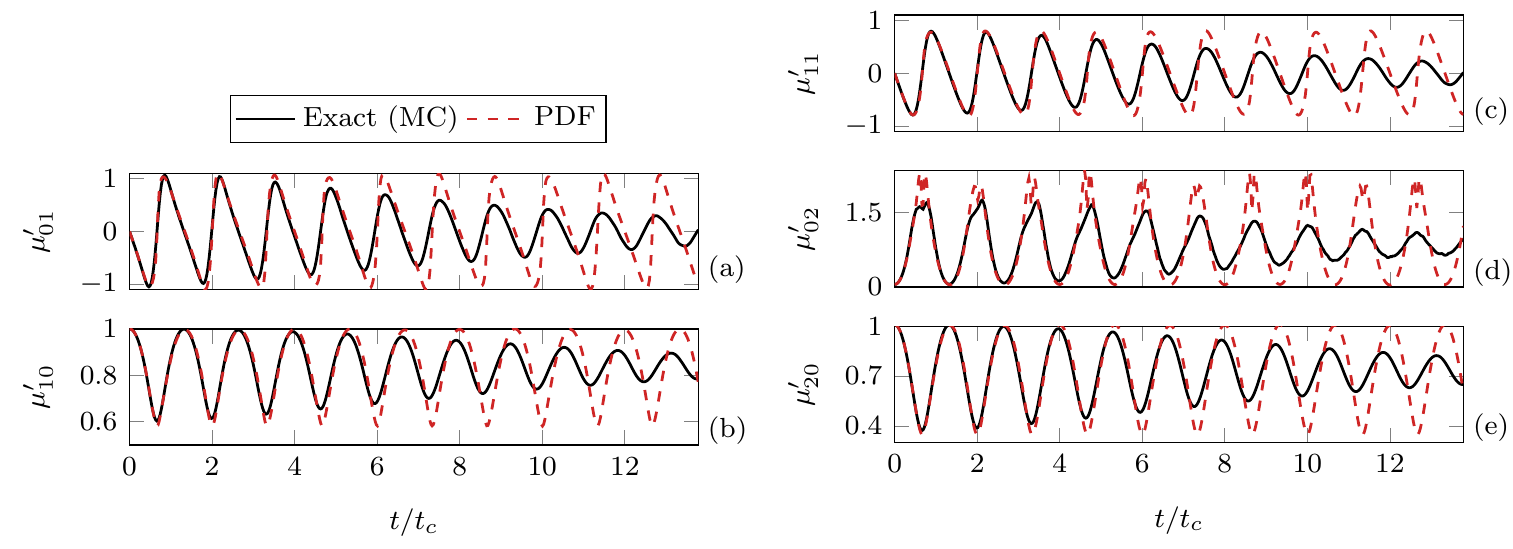}
    \caption{
        Low-order bubble population moments (a)--(e) for example case $p_o / p_\infty = 0.3$ using the PDF-based model (PDF) and Monte Carlo simulation (Exact).
        The second-order moments are normalized by their $t=0$ values and $t_c$ is the nominal collapse time.
    }
    \label{f:evolve}
\end{figure}

Computing the high-order moments associated with phase-averaged bubbly flow models requires evaluating the low-order (first- and second-order) moments.
Figure~\ref{f:evolve} shows these low-order moments for an example case over 10 periods of the mean bubble dynamics.
For both the exact and PDF-based model the moments associated with population variance, $\mom_{02}$ and $\mom_{20}$, grow and decay significantly from period-to-period.
The covariance moment $\mom_{11}$ oscillates between values near $\pm 1$, indicating correlation between the random variables.
The exact moments damp from period-to-period, whereas the moments of the PDF-based model are approximately periodic and do not display this behavior.
Thus, for this low pressure-ratio case the PDF-based model cannot accurately represent the actual statistics.

\begin{figure}[H]
    \centering
    \includegraphics[scale=1]{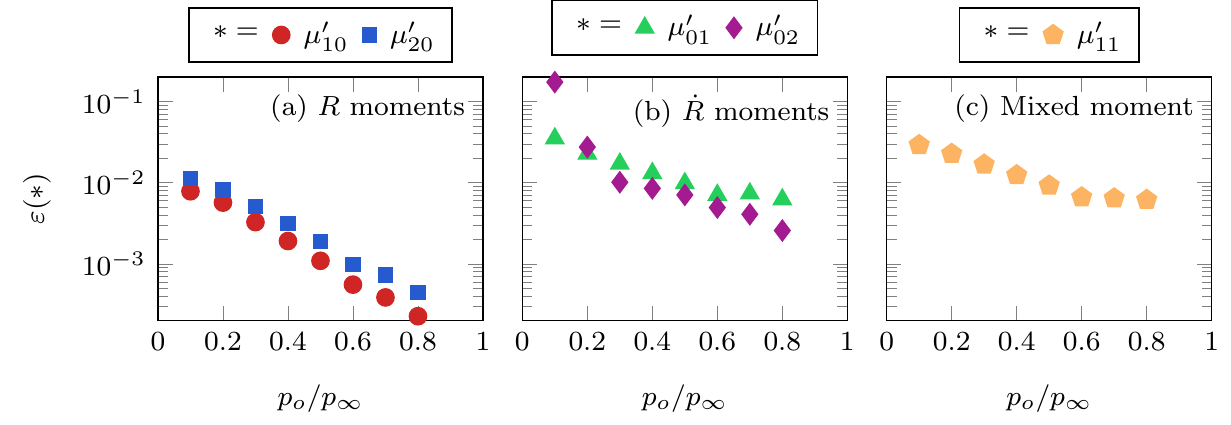}
    \caption{
        Model error $\eps$ for the low-order moments (a)--(c) over ten cycles of the mean bubble dynamics for varying $p_o/p_\infty$.
    }
    \label{f:error}
\end{figure}

Figure~\ref{f:error} shows the model error for a range of pressure ratios.
The errors of all the low-order moments increases with decreasing $p_o/p_\infty$.
The errors associated with the bubble velocity moments $\mom_{0*}$ are largest, which appears to result from the large variations that these moments have for low pressure ratios.
For pressure ratios near unity the dynamics are approximately linear and the errors are small.

\begin{figure}[H]
    \centering
    \includegraphics[scale=1]{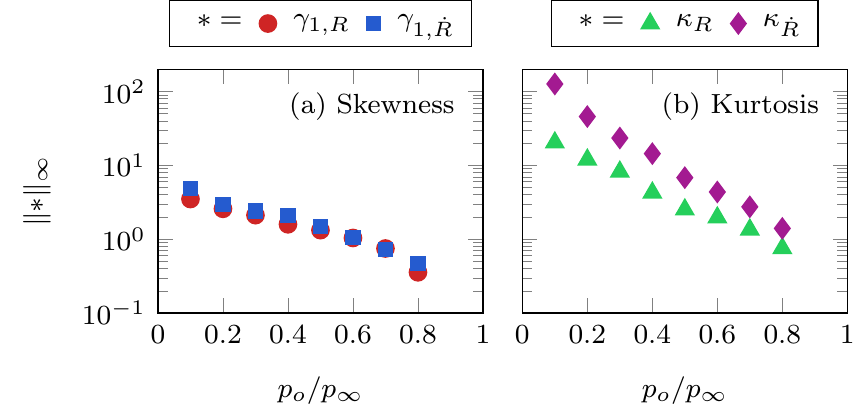}
    \caption{
        (a) Maximum Pearson's moment coefficient of skewness $\gamma_1$ and (b) excess kurtosis $\kappa$ over ten cycles of the mean bubble motion for varying pressure ratio. 
    }
    \label{f:MC}
\end{figure}

The normality of the evolving bubble dynamics quantifies the validity (or lack thereof) of the Gaussian PDF used.
Figure~\ref{f:MC} shows two high-order moments associated with non-Gaussian statistics: the maximum skewness (third standardized moment, $\gamma_1$) and excess kurtosis (fourth standardized moment, $\kappa$).
We compute these using Monte Carlo simulations.
For $p_o / p_\infty \to 1$ the dynamics are nearly linear and $\gamma_1$ and $\kappa$ are both small (less than about unity), as expected.
However, both skewness and kurtosis become large for smaller $p_o/p_\infty$.
For example, $\kappa_{\Rdot} = 126.7$ and $\gamma_{1,\Rdot} = 4.9$ for $p_o/p_\infty = 0.1$.
The large skewness results from slower bubble growth than collapse, so bubbles on spend more time at large radius and small radial velocity.
Thus, the PDF-based model, when equipped with Gaussian closure, cannot accurately predict the low-order moments for small pressure ratios.

\subsection{Higher-order moment prediction for phase-averaged models}\label{s:highorder}

For ensemble-averaged simulations, the moments required are not the usual means and variances, but instead are higher-order functions of the random variables.
Following~\citet{bryngelson19}, these are $\mom_{3(1-\gamma)0}$, $\mom_{30}$, $\mom_{21}$, and $\mom_{32}$.
Since $P$ is a multivariate Gaussian the integer moments are expressed in terms of the low-order moments as
\begin{align}
    \mom_{30} &=  3 \mom_{10} \mom_{20} -2 \mu_{10}^{\prime\,3} \label{e:highorder1}, \\
    \mom_{21} &= \mom_{01}\mom_{20} + 2 \mom_{10}\mom_{11} - 2 \mu_{10}^{\prime\,2} \mom_{01}, \\
    \mom_{32} &= \mu_{10}^{\prime\,3} \left( 6 \mu_{01}^{\prime\,2} - 2 \mom_{02} \right)  - 
                    12 \mu_{10}^{\prime\,2} \mom_{01} \mom_{11} + 
                    6  \mom_{01} \mom_{20} \mom_{11} +
                    \mom_{10} 
                    \left[ 3 \mom_{20} 
                        \left( \mom_{02} - 2 \mu_{01}^{\prime\,2} \right) + 6 \mu_{11}^{\prime\,2} 
                    \right]
    \label{e:highorder3}
\end{align}
Adaptive Gaussian quadrature computes the non-integer moment $\mom_{3(1-\gamma)0}$ via~\eqref{e:mom}.

\begin{figure}
	\centering
	\includegraphics[scale=1]{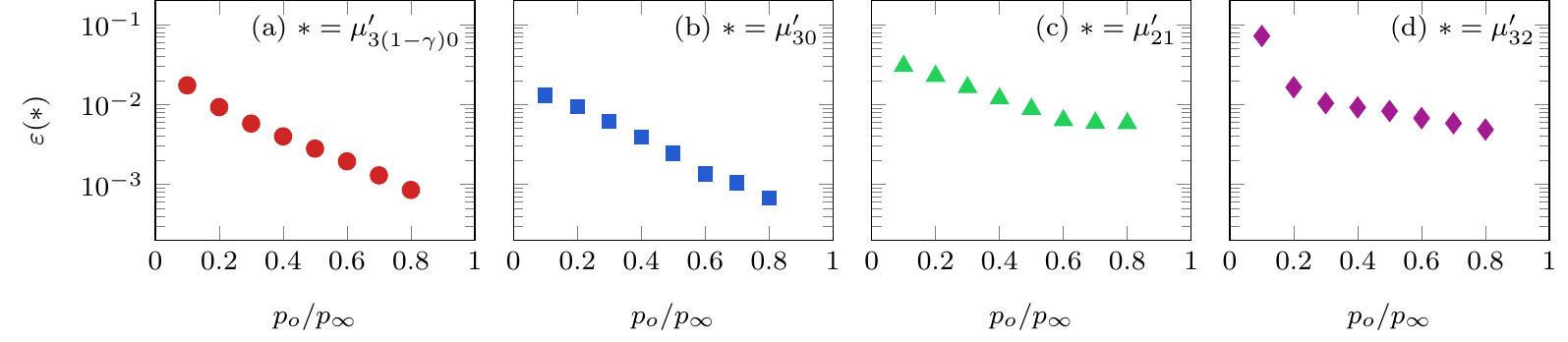}
    \caption{PDF-basd model error $\eps$ associated with specific distribution moments (a)--(d).}
	\label{f:enserror}
\end{figure}

Figure~\ref{f:enserror} shows the relative model error of the phase-averaged model moments.
Similar to figure~\ref{f:error}, the errors grow with decreasing pressure ratio and the moments associated with the bubble radius $R$ have smaller errors, including the non-integer moment $\mom_{3(1-\gamma)0}$.
These errors are large for small $p_o / p_\infty$, and the highest-order moment $\mom_{32}$ has the largest error.
However, this is still comparable to the errors observed for $\mom_{02}$.

\section{Model augmentation via LSTM recurrent neural networks}\label{s:ML}

The Gaussian-closure-based method performed well for modest pressure ratios, but poorly for strong bubble dynamics because of population skewness and kurtosis.
Representing such high-order statistics via a more general density function is challenging because of the moment problem discussed in section~\ref{s:intro}. 
Instead, we will employ a machine learning formulation to complement the Gaussian closure method. 
This approach can effectively augment imperfect dynamical systems (e.g.\ \citep{wan18plos}).
Here, it attempts to improve prediction of both low-order moment evolution and high-order moment evaluation. 
The machine learning component is an LSTM RNN, which allows incorporation of memory effects in the resulting machine-learned equations.
Thus, the moment system \eqref{m_system} has non-time-local closures and is non-Markovian~\citep{wan18}

\subsection{Low-order moment prediction}\label{s:ML_loworder}

Using the Gaussian closure of section~\ref{s:pdf} as a starting point, a machine-learned forcing term $\vbf^\mathrm{ML}$ augments the low-order moment evolution~\eqref{e:system} as
\begin{gather}
    \frac{\partial \bmom}{\partial t} = \vbf(\bmom) + \vbf^\mathrm{ML}(\bmom).
    \label{m_system}
\end{gather}
A separate single-layer LSTM RNN (each with 32 time delays) determines each component of this term.
The Monte Carlo time history of $\bmom$ for cases $p_o / p_\infty = \{ 0.15, 0.25, \dots, 0.85 \}$ trains the neural networks and provide the first 32 time delays. 

\begin{figure}
    \centering
    \includegraphics[scale=1]{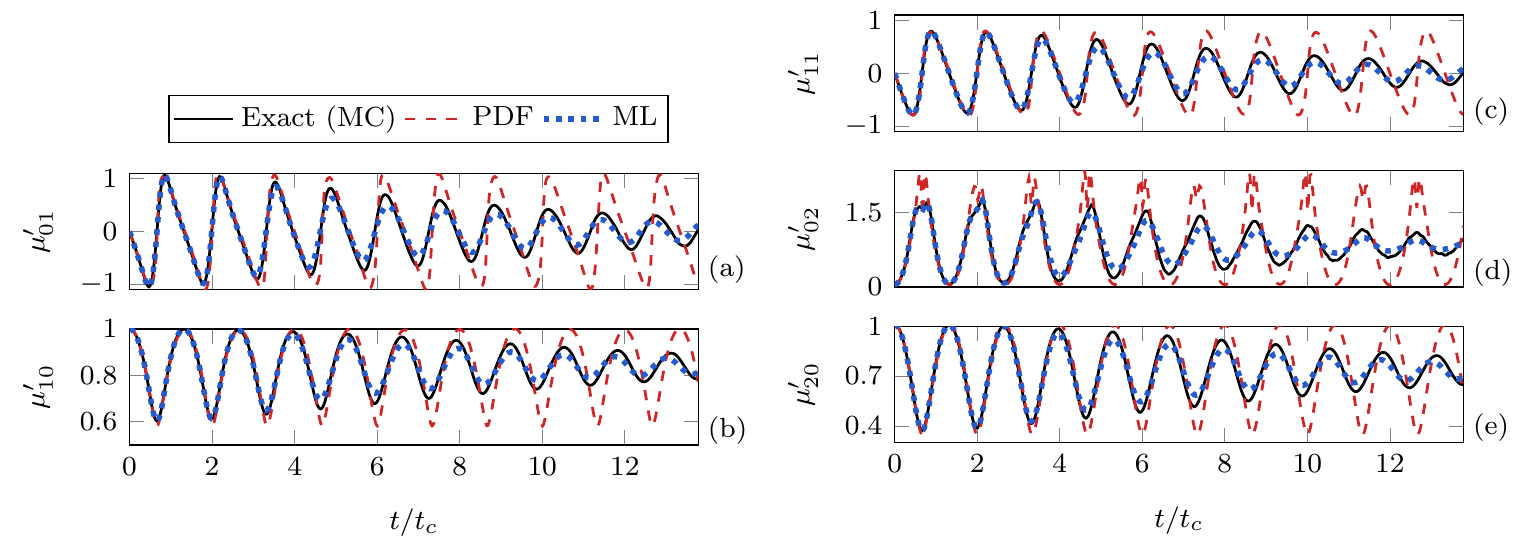}
    \caption{
        Low-order bubble population moments (a)--(e) for example case $p_o / p_\infty = 0.3$ using the PDF-based model (PDF), the neural-network-augmented model (ML), and Monte Carlo simulation (exact).
        The second-order moments are normalized by their $t=0$ values and $t_c$ is the nominal collapse time.
    }
    \label{f:coupled}
\end{figure}

Figure~\ref{f:coupled} shows these low-order moments for the neural-network-augmented model.
Even for this relatively low pressure-ratio case, the moments associated with machine learning approach are much closer to the exact data than the PDF-based model alone.
This including the relative damping of all moments, which the PDF-based model could not represent.

\begin{figure}
    \centering
    \includegraphics[scale=1]{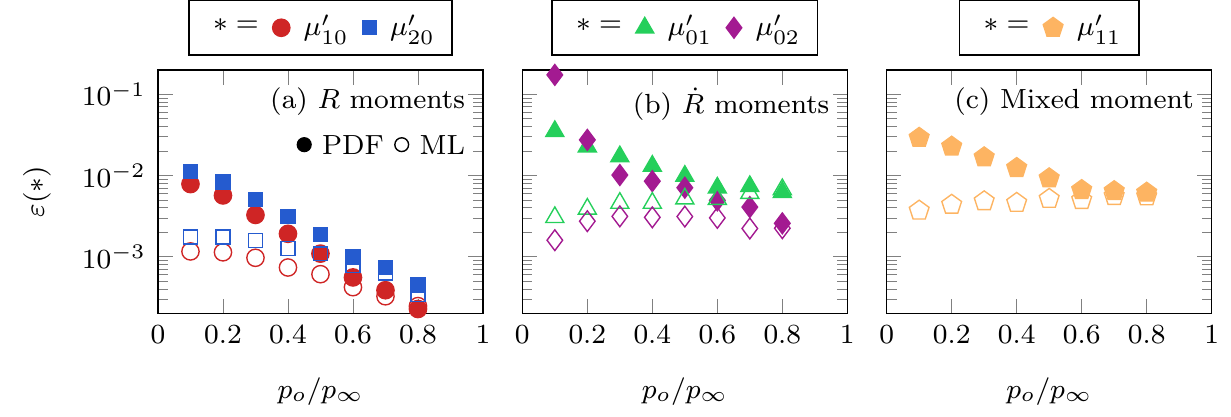}
    \caption{
        Model error $\eps$ for the low-order moments (a)--(c) for the PDF-based model (PDF) and ML-augmented PDF-based model (ML) at varying pressure ratio.
    }
    \label{f:coupled_loworder}
\end{figure}

Figure~\ref{f:coupled_loworder} shows the error of the PDF-based model and its augmentation via neural networks.
The machine learning approach significantly decreases the model error for all moments for $p_o/p_\infty \lesssim 0.5$, while the errors for larger pressure ratios only decrease modestly.
These errors are approximately the same for the bubble radius and radial velocity moments.
Thus, the $\Rdot$ moment predictions improve the most, since they had the largest errors for PDF-based model alone.
For example, for $p_o/p_\infty = 0.2$ the ML error is only $8\%$ of the PDF error for $\mom_{01}$ and $0.9\%$ of it for $\mom_{02}$. 
Note that for the lowest pressure ratio we consider, $p_o / p_\infty = 0.1$, including the $\bff$ term associated with the Gaussian closure did not improve our results. 
This is because this case has significant non-Gaussian features.
Thus, for this case we trained a neural network on the data itself without $\bff$.

\subsection{Higher-order moment prediction for phase-averaged models}\label{s:ML_highorder}

\begin{figure}
    \centering
	\includegraphics[scale=1]{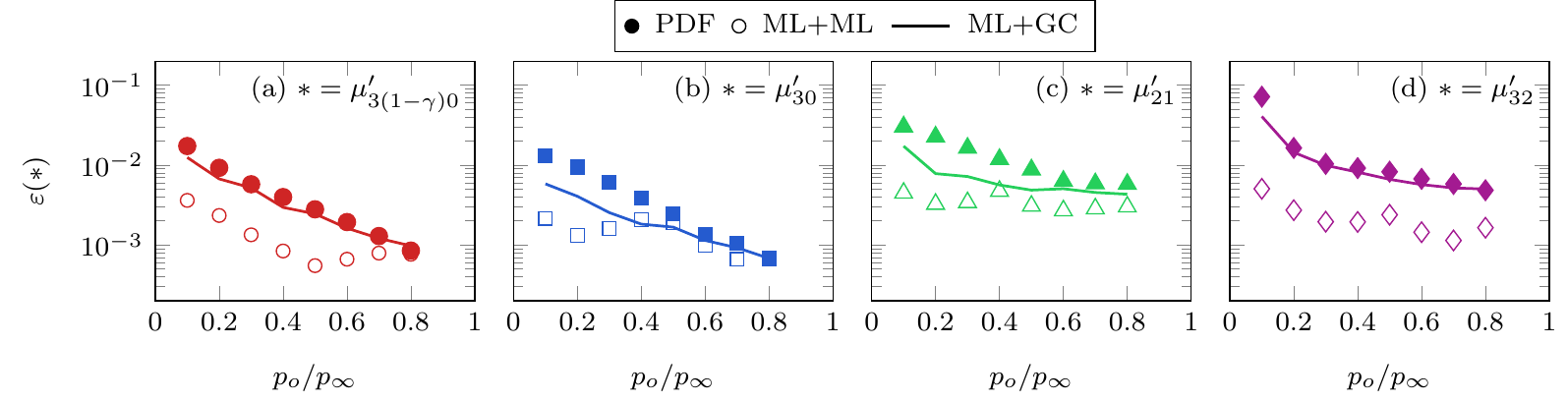}
    \caption{Model error $\eps$ associated with specific distribution moments (a)--(d) for the PDF-based model (PDF), the ML-augmented PDF-based model (ML+GC), and the ML-augmented PDF-based model, augmented with an additional LSTM RNN for these moments (ML+ML).
    }
    \label{f:coupled_highorder}
\end{figure}

This section analyzes the higher-order moments of section~\ref{s:highorder} for the improved model predictions.
Figure~\ref{f:coupled_highorder} shows the errors associated with these moments for the augmented PDF-based model.
The PDF-based model errors (PDF) are also those of figure~\ref{f:enserror} and ML-augmented errors (ML+GC) follow from assuming a Gaussian PDF for the higher-order moments using the low-order moments of section~\ref{s:ML_loworder}. 
We see that this approach alone reduces the error from the PDF-only model for the $\mom_{30}$ and $\mom_{21}$ moments. 
However, the non-integer moment errors do not decrease significantly.
This is because the primary error results from non-Gaussian statistics, and so assuming Gaussianity for the other moments precludes accurate prediction.

An additional LSTM neural network with output $\vbg^\mathrm{ML}$ is used to reduce these errors as
\begin{gather}
    \bmom_{\mathrm{ML}} = \bmom_\mathrm{HG}(\bmom) + \vbg^\mathrm{ML}(\bmom),
\end{gather}
where $\bmom_\mathrm{HG}$ is the column vector of high-order moments as approximated via Gaussian statistics following \eqref{e:highorder1}--\eqref{e:highorder3} and $\bmom_\mathrm{ML}$ are the new predictions (labeled as ML+ML in figure~\ref{f:coupled_highorder}).
The low-order moments $\bmom$ and the residual of the truth-value of the higher-order moments (computed via Monte Carlo data) and $\bmom_\mathrm{HG}$ train this neural network.
Figure~\ref{f:coupled_highorder} shows these results (ML+ML) for verification (out-of-training-set) pressure ratios.
This approach reduces $\eps$ from the ML+GC and PDF results for the $\mom_{3(1-\gamma)0}$ and $\mom_{32}$ moments and reduces it further from the ML+GC results for the other moments.
For example, for $p_o/p_\infty = 0.1$ the ML+ML error is only $7\%$ of the PDF error for the $\mom_{32}$ moment and $20\%$ of it for $\mom_{3(1-\gamma)0}$. 

\section{Discussion and outlook}\label{s:limit}

The kernel of the integrals of~\eqref{e:rhs} is $P$, which is ill-posed for $\rho \to 1$ and fixed independent coordinates.
In practice the model only displayed this issue when the initial conditions require $\rho = 1$, as the moment system did not generate such strong correlations otherwise.
Still, evaluating the $\rho(t=0) = 1$ case requires a coordinate transformation, such as that of~\citet{glazunov12}.

We did not analyze the cost of evaluating the integrals of~\eqref{e:rhs} above.
However, obtaining less than $1\%$ relative error requires only about 20 integrand evaluations when using adaptive Gauss quadrature for the $p_o / p_\infty = 0.3$ case of section~\ref{s:nonlinear} at $t = t_c$.
This compares favorably with other moment methods because only five moments are evolved (for the cases of section~\ref{s:nonlinear}).
This is because evolving more degrees of freedom in a simulation environment is expensive, particularly for the high-order interface-capturing methods often used for bubbly flows~\citep{coralic14}.
Of course, there are other ways to evaluate these integrals.
For example, interpolated look-up tables are an efficient treatment for problems of this type: molecular dynamic simulations often use these for particle-pair potentials~\citep{wolff99,rapaport04} and associated integral quantities~\citep{stave90}, chemical-reacting system simulations use them for reduced-order chemistry~\citep{pope97} and flame models~\citep{jha12}.
Such a method would be useful because many parts of the flow fields are likely to see similar conditions at any instance in time.
Another route is accelerating their evaluation via neural networks.
Indeed, this has been used to evaluate integrals corresponding to combustion systems~\citep{blasco98}.

Actual simulation environments have spatial inhomogeneities that can lead to numerical instabilities of the quadrature weights for QBMMs because of unrealizable quadratures~\citep{wright07,mazzei10,mazzei12}.
This is a complex problem that the method presented here did not address directly.
However, this method does not require fixed quadrature nodes that QBMMs use and so it is unclear whether and how such instabilities could develop. 

\section{Conclusions}\label{s:conclusion}

A moment method for predicting the statistics of a population of dilute, cavitating bubbles was presented.  The moment equations are closed via a Gaussian probability density function, and only require evolution of the first two moments.  
In order to correct for errors incurred in the closure, it is augmented by a recurrent neural network.
This data-driven representation was trained on Monte Carlo data to correct the low-order moments, substantially improving predictions.
For example, for low pressure ratio $p_o/p_\infty = 0.2$ the ML-augmented model error was only $0.9\%$ of the unaugmented method for the $\mom_{02}$ moment, which had the largest error without the neural network.

The higher-order moments required to close phase-averaged bubbly flow models cannot, however, be predicted based on the (corrected) low-order moments, since they will contain errors associated with non-Gaussian statistics.
Using an additional neural network, trained on only Monte Carlo and low-order moment data, prediction of these high-order moments improved significantly.
For example, for the lowest pressure ratio case $p_o/p_\infty = 0.1$ the error was only $7\%$ of the PDF error for the highest-order moment.

These results suggest that RNN-augmented moment models can efficiently evaluate the bubbly flow statistics required to close phase-averaged models.  
Future work will evaluate the performance of these models in coupled bubbly-flow simulations.

\section*{Acknowledgments}

The US Office of Naval Research supported this work under MURI grant N0014-17-1-2676.

\section*{References}
\bibliographystyle{elsarticle-num-names}
\bibliography{main}

\begin{thebibliography}{44}
\expandafter\ifx\csname natexlab\endcsname\relax\def\natexlab#1{#1}\fi
\providecommand{\url}[1]{\texttt{#1}}
\providecommand{\href}[2]{#2}
\providecommand{\path}[1]{#1}
\providecommand{\DOIprefix}{doi:}
\providecommand{\ArXivprefix}{arXiv:}
\providecommand{\URLprefix}{URL: }
\providecommand{\Pubmedprefix}{pmid:}
\providecommand{\doi}[1]{\href{http://dx.doi.org/#1}{\path{#1}}}
\providecommand{\Pubmed}[1]{\href{pmid:#1}{\path{#1}}}
\providecommand{\bibinfo}[2]{#2}
\ifx\xfnm\relax \def\xfnm[#1]{\unskip,\space#1}\fi
\bibitem[{Akhiezer(1965)}]{akhiezer65}
\bibinfo{author}{N.~I. Akhiezer}, \bibinfo{title}{The classical moment problem:
  and some related questions in analysis}, University mathematical monographs,
  \bibinfo{publisher}{{Oliver \& Boyd}}, \bibinfo{year}{1965}.
\bibitem[{Ando(2010)}]{ando10}
\bibinfo{author}{K.~Ando}, \bibinfo{title}{Effects of polydispersity in bubbly
  flows}, Ph.D. thesis, California Institute of Technology,
  \bibinfo{year}{2010}.
\bibitem[{Ando et~al.(2011)Ando, Colonius, and Brennen}]{ando11}
\bibinfo{author}{K.~Ando}, \bibinfo{author}{T.~Colonius},
  \bibinfo{author}{C.~E. Brennen},
\newblock \bibinfo{title}{Numerical simulation of shock propagation in a
  polydisperse bubbly liquid},
\newblock \bibinfo{journal}{Int. J. Mult. Flow} \bibinfo{volume}{37}
  (\bibinfo{year}{2011}) \bibinfo{pages}{596--608}.
\bibitem[{Blasco et~al.(1998)Blasco, Fueyo, Dopazo, and Ballester}]{blasco98}
\bibinfo{author}{J.~A. Blasco}, \bibinfo{author}{N.~Fueyo},
  \bibinfo{author}{C.~Dopazo}, \bibinfo{author}{J.~Ballester},
\newblock \bibinfo{title}{Modelling the temporal evolution of a reduced
  combustion chemical system with an artificial neural network},
\newblock \bibinfo{journal}{Combust. Flame} \bibinfo{volume}{113}
  (\bibinfo{year}{1998}) \bibinfo{pages}{38--52}.
\bibitem[{Brennen(2015)}]{brennen15}
\bibinfo{author}{C.~E. Brennen},
\newblock \bibinfo{title}{Cavitation in medicine},
\newblock \bibinfo{journal}{Interface Focus} \bibinfo{volume}{5}
  (\bibinfo{year}{2015}).
\bibitem[{Bryngelson et~al.(2019)Bryngelson, Schmidmayer, and
  Colonius}]{bryngelson19}
\bibinfo{author}{S.~H. Bryngelson}, \bibinfo{author}{K.~Schmidmayer},
  \bibinfo{author}{T.~Colonius},
\newblock \bibinfo{title}{A quantitative comparison of phase-averaged models
  for bubbly, cavitating flows},
\newblock \bibinfo{journal}{Int. J. Mult. Flow} \bibinfo{volume}{115}
  (\bibinfo{year}{2019}) \bibinfo{pages}{137--143}.
\bibitem[{Capecelatro and Desjardins(2013)}]{capecelatro13}
\bibinfo{author}{J.~Capecelatro}, \bibinfo{author}{O.~Desjardins},
\newblock \bibinfo{title}{An {E}uler--{L}agrange strategy for simulating
  particle-laden flows},
\newblock \bibinfo{journal}{J. Comp. Phys.} \bibinfo{volume}{238}
  (\bibinfo{year}{2013}) \bibinfo{pages}{1--31}.
\bibitem[{Chalons et~al.(2012)Chalons, Kah, and Massot}]{chalons12}
\bibinfo{author}{C.~Chalons}, \bibinfo{author}{D.~Kah},
  \bibinfo{author}{M.~Massot},
\newblock \bibinfo{title}{Beyond pressureless gas dynamics: {Q}uadrature-based
  velocity moment models},
\newblock \bibinfo{journal}{Comm. Math. Sci.} \bibinfo{volume}{10}
  (\bibinfo{year}{2012}) \bibinfo{pages}{1241--1272}.
\bibitem[{Chang et~al.(2008)Chang, Ebert, Young, Liu, Mahesh, Jang, and
  Shearer}]{chang08}
\bibinfo{author}{P.~A. Chang}, \bibinfo{author}{M.~Ebert},
  \bibinfo{author}{Y.~L. Young}, \bibinfo{author}{Z.~Liu},
  \bibinfo{author}{K.~Mahesh}, \bibinfo{author}{H.~Jang},
  \bibinfo{author}{M.~Shearer},
\newblock \bibinfo{title}{Propeller forces and structural response due to
  crashback},
\newblock in: \bibinfo{booktitle}{27th Symposium on Naval Hydrodynamics},
  \bibinfo{year}{2008}.
\bibitem[{Colonius et~al.(2008)Colonius, Hagmeijer, Ando, and
  Brennen}]{colonius08}
\bibinfo{author}{T.~Colonius}, \bibinfo{author}{R.~Hagmeijer},
  \bibinfo{author}{K.~Ando}, \bibinfo{author}{C.~E. Brennen},
\newblock \bibinfo{title}{Statistical equilibrium of bubble oscillations in
  dilute bubbly flows},
\newblock \bibinfo{journal}{Phys. Fluids} \bibinfo{volume}{20}
  (\bibinfo{year}{2008}).
\bibitem[{Coralic and Colonius(2006)}]{coralic14}
\bibinfo{author}{V.~Coralic}, \bibinfo{author}{T.~Colonius},
\newblock \bibinfo{title}{Finite-volume {WENO} scheme for viscous compressible
  multicomponent flow problems},
\newblock \bibinfo{journal}{J. Comp. Phys.} \bibinfo{volume}{219}
  (\bibinfo{year}{2006}) \bibinfo{pages}{715--732}.
\bibitem[{Desjardins et~al.(2008)Desjardins, Fox, and Villedieu}]{desjardins08}
\bibinfo{author}{O.~Desjardins}, \bibinfo{author}{R.~O. Fox},
  \bibinfo{author}{P.~Villedieu},
\newblock \bibinfo{title}{A quadrature-based moment method for dilute
  fluid-particle flows},
\newblock \bibinfo{journal}{J. Comp. Phys.} \bibinfo{volume}{227}
  (\bibinfo{year}{2008}) \bibinfo{pages}{2514--2539}.
\bibitem[{Fox(2009)}]{fox09}
\bibinfo{author}{R.~O. Fox},
\newblock \bibinfo{title}{Higher-order quadrature-based moment methods for
  kinetic equations},
\newblock \bibinfo{journal}{J. Comp. Phys.} \bibinfo{volume}{228}
  (\bibinfo{year}{2009}) \bibinfo{pages}{7771--7791}.
\bibitem[{Glazunov and Zhang(2012)}]{glazunov12}
\bibinfo{author}{A.~A. Glazunov}, \bibinfo{author}{J.~Zhang},
\newblock \bibinfo{title}{A note on the bivariate distribution representation
  of two perfectly correlated random variables by {D}irac's $\delta$-function},
\newblock \bibinfo{journal}{arXiv preprint arXiv:1205.0933}
  (\bibinfo{year}{2012}).
\bibitem[{Heylmun et~al.(2019)Heylmun, Kong, Passalacqua, and Fox}]{heylmun19}
\bibinfo{author}{J.~C. Heylmun}, \bibinfo{author}{B.~Kong},
  \bibinfo{author}{A.~Passalacqua}, \bibinfo{author}{R.~Fox},
\newblock \bibinfo{title}{A quadrature-based moment method for polydisperse
  bubbly flows},
\newblock \bibinfo{journal}{Comp. Phys. Comm.}  (\bibinfo{year}{2019}).
\bibitem[{Hochreiter and Schmidhuber(1997)}]{hochreiter1997long}
\bibinfo{author}{S.~Hochreiter}, \bibinfo{author}{J.~Schmidhuber},
\newblock \bibinfo{title}{Long short-term memory},
\newblock \bibinfo{journal}{Neural Comput.} \bibinfo{volume}{9}
  (\bibinfo{year}{1997}) \bibinfo{pages}{1735--1780}.
\bibitem[{Hulburt and Katz(1964)}]{hulburt64}
\bibinfo{author}{H.~M. Hulburt}, \bibinfo{author}{S.~Katz},
\newblock \bibinfo{title}{Some problems in particle technology. {A} statistical
  mechanical formulation},
\newblock \bibinfo{journal}{Chem. Eng. Sci.} \bibinfo{volume}{19}
  (\bibinfo{year}{1964}) \bibinfo{pages}{555--574}.
\bibitem[{Ikeda et~al.(2006)Ikeda, Yoshizawa, Masakata, Allen, Takagi, Ohta,
  Kitamura, and Matsumoto}]{ikeda06}
\bibinfo{author}{T.~Ikeda}, \bibinfo{author}{S.~Yoshizawa},
  \bibinfo{author}{T.~Masakata}, \bibinfo{author}{J.~S. Allen},
  \bibinfo{author}{S.~Takagi}, \bibinfo{author}{N.~Ohta},
  \bibinfo{author}{T.~Kitamura}, \bibinfo{author}{Y.~Matsumoto},
\newblock \bibinfo{title}{Cloud cavitation control for lithotripsy using high
  intensity focused ultrasound},
\newblock \bibinfo{journal}{Ultrasound Med. Biol.} \bibinfo{volume}{32}
  (\bibinfo{year}{2006}) \bibinfo{pages}{1383--1397}.
\bibitem[{Jha and Groth(2012)}]{jha12}
\bibinfo{author}{P.~K. Jha}, \bibinfo{author}{C.~P.~T. Groth},
\newblock \bibinfo{title}{Tabulated chemistry approaches for laminar flames:
  {E}valuation of flame-prolongation of {ILDM} and flamelet methods},
\newblock \bibinfo{journal}{Combust. Theor. Model.} \bibinfo{volume}{16}
  (\bibinfo{year}{2012}) \bibinfo{pages}{31--57}.
\bibitem[{Kong and Fox(2019)}]{kong19}
\bibinfo{author}{B.~Kong}, \bibinfo{author}{R.~O. Fox},
\newblock \bibinfo{title}{A moment-based kinetic theory model for polydisperse
  gas--particle flows},
\newblock \bibinfo{journal}{Powder Technol.}  (\bibinfo{year}{2019}).
\bibitem[{Laksari et~al.(2015)Laksari, Assari, Seibold, Sadeghipour, and
  Darvish}]{laksari15}
\bibinfo{author}{K.~Laksari}, \bibinfo{author}{S.~Assari},
  \bibinfo{author}{B.~Seibold}, \bibinfo{author}{K.~Sadeghipour},
  \bibinfo{author}{K.~Darvish},
\newblock \bibinfo{title}{Computational simulation of the mechanical response
  of brain tissue under blast loading},
\newblock \bibinfo{journal}{Biomech. Model. Mechanobiol.} \bibinfo{volume}{14}
  (\bibinfo{year}{2015}) \bibinfo{pages}{459--472}.
\bibitem[{Maeda and Colonius(2019)}]{maeda19}
\bibinfo{author}{K.~Maeda}, \bibinfo{author}{T.~Colonius},
\newblock \bibinfo{title}{Bubble cloud dynamics in an ultrasound field},
\newblock \bibinfo{journal}{J. Fluid Mech.} \bibinfo{volume}{862}
  (\bibinfo{year}{2019}) \bibinfo{pages}{1105--1134}.
\bibitem[{Marchisio and Fox(2005)}]{marchisio05}
\bibinfo{author}{D.~L. Marchisio}, \bibinfo{author}{R.~O. Fox},
\newblock \bibinfo{title}{Solution of population balance equations using the
  direct quadrature method of moments},
\newblock \bibinfo{journal}{J. Aerosol Sci.} \bibinfo{volume}{36}
  (\bibinfo{year}{2005}) \bibinfo{pages}{43--73}.
\bibitem[{Mazzei et~al.(2010)Mazzei, Marchisio, and Lettieri}]{mazzei10}
\bibinfo{author}{L.~Mazzei}, \bibinfo{author}{D.~Marchisio},
  \bibinfo{author}{P.~Lettieri},
\newblock \bibinfo{title}{Direct quadrature method of moments for the mixing of
  inert polydisperse fluidized powders and the role of numerical diffusion},
\newblock \bibinfo{journal}{Ind. Eng. Chem. Res.} \bibinfo{volume}{49}
  (\bibinfo{year}{2010}) \bibinfo{pages}{5141--5152}.
\bibitem[{Mazzei et~al.(2012)Mazzei, Marchisio, and Lettieri}]{mazzei12}
\bibinfo{author}{L.~Mazzei}, \bibinfo{author}{D.~L. Marchisio},
  \bibinfo{author}{P.~Lettieri},
\newblock \bibinfo{title}{A new quadrature-based moment method for the mixing
  of inert polydisperse fluidized powders in commercial {CFD} codes},
\newblock \bibinfo{journal}{AIChE J.} \bibinfo{volume}{58}
  (\bibinfo{year}{2012}) \bibinfo{pages}{3054--3069}.
\bibitem[{McGraw(1997)}]{mcgraw97}
\bibinfo{author}{R.~McGraw},
\newblock \bibinfo{title}{Description of aerosol dynamics by the quadrature
  method of moments},
\newblock \bibinfo{journal}{Aerosol Sci. Technol.} \bibinfo{volume}{27}
  (\bibinfo{year}{1997}) \bibinfo{pages}{255--265}.
\bibitem[{Moyal(1949)}]{moyal49}
\bibinfo{author}{J.~E. Moyal},
\newblock \bibinfo{title}{Stochastic processes and statistical physics},
\newblock \bibinfo{journal}{J. Roy. Stat. Soc. B} \bibinfo{volume}{11}
  (\bibinfo{year}{1949}).
\bibitem[{Pishchalnikov et~al.(2003)Pishchalnikov, Sapozhnikov, Bailey,
  Williams, Cleveland, Colonius, Crum, Evan, and McAteer}]{pishchalnikov03}
\bibinfo{author}{Y.~A. Pishchalnikov}, \bibinfo{author}{O.~A. Sapozhnikov},
  \bibinfo{author}{M.~R. Bailey}, \bibinfo{author}{J.~C. Williams},
  \bibinfo{author}{R.~O. Cleveland}, \bibinfo{author}{T.~Colonius},
  \bibinfo{author}{L.~A. Crum}, \bibinfo{author}{A.~P. Evan},
  \bibinfo{author}{J.~A. McAteer},
\newblock \bibinfo{title}{Cavitation bubble cluster activity in the breakage of
  kidney stones by lithotripter shockwaves},
\newblock \bibinfo{journal}{J. Endourol.} \bibinfo{volume}{17}
  (\bibinfo{year}{2003}) \bibinfo{pages}{435--446}.
\bibitem[{Pope(1997)}]{pope97}
\bibinfo{author}{S.~B. Pope},
\newblock \bibinfo{title}{Computationally efficient implementation of
  combustion chemistry using in situ adaptive tabulation},
\newblock \bibinfo{journal}{Combust. Theor. Model.} \bibinfo{volume}{1}
  (\bibinfo{year}{1997}) \bibinfo{pages}{41--63}.
\bibitem[{Rapaport(2004)}]{rapaport04}
\bibinfo{author}{D.~C. Rapaport}, \bibinfo{title}{The Art of Molecular Dynamics
  Simulation}, \bibinfo{edition}{2} ed., \bibinfo{publisher}{Cambridge
  University Press}, \bibinfo{year}{2004}.
\bibitem[{Shimada et~al.(2000)Shimada, Matsumoto, and Kobayashi}]{shimada00}
\bibinfo{author}{M.~Shimada}, \bibinfo{author}{Y.~Matsumoto},
  \bibinfo{author}{T.~Kobayashi},
\newblock \bibinfo{title}{Influence of the nuclei size distribution on the
  collapsing behavior of the cloud cavitation},
\newblock \bibinfo{journal}{JSME Int. J. Ser. B} \bibinfo{volume}{43}
  (\bibinfo{year}{2000}) \bibinfo{pages}{380--385}.
\bibitem[{Smereka(2002)}]{smereka02}
\bibinfo{author}{P.~Smereka},
\newblock \bibinfo{title}{A {V}lasov equation for pressure wave propagation in
  bubbly fluids},
\newblock \bibinfo{journal}{J. Fluid Mech.} \bibinfo{volume}{454}
  (\bibinfo{year}{2002}) \bibinfo{pages}{287--325}.
\bibitem[{Srinivasan et~al.(2019)Srinivasan, Guastoni, Azizpour, Schlatter, and
  Vinuesa}]{srinivasan19}
\bibinfo{author}{P.~A. Srinivasan}, \bibinfo{author}{L.~Guastoni},
  \bibinfo{author}{H.~Azizpour}, \bibinfo{author}{P.~Schlatter},
  \bibinfo{author}{R.~Vinuesa},
\newblock \bibinfo{title}{Predictions of turbulent shear flows using deep
  neural networks},
\newblock \bibinfo{journal}{Phys. Rev. Fluids} \bibinfo{volume}{4}
  (\bibinfo{year}{2019}) \bibinfo{pages}{054603}.
\bibitem[{Stave et~al.(1990)Stave, Sanders, Raeker, and DePristo}]{stave90}
\bibinfo{author}{M.~S. Stave}, \bibinfo{author}{D.~E. Sanders},
  \bibinfo{author}{T.~J. Raeker}, \bibinfo{author}{A.~E. DePristo},
\newblock \bibinfo{title}{Corrected effective medium method. {V}.
  {S}implifications for molecular dynamics and {M}onte {C}arlo simulations},
\newblock \bibinfo{journal}{J. Chem. Phys.} \bibinfo{volume}{93}
  (\bibinfo{year}{1990}) \bibinfo{pages}{4413--4426}.
\bibitem[{Stieltjes(1894)}]{stieltjes94}
\bibinfo{author}{T.-J. Stieltjes},
\newblock \bibinfo{title}{Recherches sur les fractions continues},
\newblock in: \bibinfo{booktitle}{Annales de la Facult{\'e} des sciences de
  Toulouse: Math{\'e}matiques}, volume~\bibinfo{volume}{8},
  \bibinfo{year}{1894}, pp. \bibinfo{pages}{J1--J122}.
\bibitem[{Vanni(2000)}]{vanni00}
\bibinfo{author}{M.~Vanni},
\newblock \bibinfo{title}{Approximate population balance equations for
  aggregation breakage processes},
\newblock \bibinfo{journal}{J. Colloid Interface Sci.} \bibinfo{volume}{221}
  (\bibinfo{year}{2000}) \bibinfo{pages}{143--160}.
\bibitem[{Wan et~al.(2018)Wan, Vlachas, Koumoutsakos, and Sapsis}]{wan18plos}
\bibinfo{author}{Z.~Wan}, \bibinfo{author}{P.~Vlachas},
  \bibinfo{author}{P.~Koumoutsakos}, \bibinfo{author}{T.~Sapsis},
\newblock \bibinfo{title}{Data-assisted reduced-order modeling of extreme
  events in complex dynamical systems},
\newblock \bibinfo{journal}{PLOS One} \bibinfo{volume}{24 May}
  (\bibinfo{year}{2018}).
\bibitem[{Wan et~al.(2019)Wan, Karnakov, Koumoutsakos, and Sapsis}]{wan19}
\bibinfo{author}{Z.~Y. Wan}, \bibinfo{author}{P.~Karnakov},
  \bibinfo{author}{P.~Koumoutsakos}, \bibinfo{author}{T.~Sapsis},
\newblock \bibinfo{title}{Bubbles in turbulent flows: {D}ata-driven, kinematic
  models with history terms},
\newblock \bibinfo{journal}{arXiv:1910.02068} \bibinfo{volume}{Submitted}
  (\bibinfo{year}{2019}).
\bibitem[{Wan and Sapsis(2018)}]{wan18}
\bibinfo{author}{Z.~Y. Wan}, \bibinfo{author}{T.~P. Sapsis},
\newblock \bibinfo{title}{Machine learning the kinematics of spherical
  particles in fluid flows},
\newblock \bibinfo{journal}{J. Fluid Mech.} \bibinfo{volume}{857}
  (\bibinfo{year}{2018}) \bibinfo{pages}{R2}.
\bibitem[{Wolff and Rudd(1999)}]{wolff99}
\bibinfo{author}{D.~Wolff}, \bibinfo{author}{W.~G. Rudd},
\newblock \bibinfo{title}{Tabulated potentials in molecular dynamics
  simulations},
\newblock \bibinfo{journal}{Comp. Phys. Comm.} \bibinfo{volume}{120}
  (\bibinfo{year}{1999}) \bibinfo{pages}{20 -- 32}.
\bibitem[{{Wright Jr.}(2007)}]{wright07}
\bibinfo{author}{D.~L. {Wright Jr.}},
\newblock \bibinfo{title}{Numerical advection of moments of the particle size
  distribution in {E}ulerian models},
\newblock \bibinfo{journal}{J. Aerosol Sci.} \bibinfo{volume}{38}
  (\bibinfo{year}{2007}) \bibinfo{pages}{352--369}.
\bibitem[{Zhang and Prosperetti(1994)}]{zhang94}
\bibinfo{author}{D.~Z. Zhang}, \bibinfo{author}{A.~Prosperetti},
\newblock \bibinfo{title}{Ensemble phase-averaged equations for bubbly flows},
\newblock \bibinfo{journal}{Phys. Fluids} \bibinfo{volume}{6}
  (\bibinfo{year}{1994}).
\bibitem[{Zhao et~al.(2007)Zhao, Maisels, Matsoukas, and Zheng}]{zhao07}
\bibinfo{author}{H.~Zhao}, \bibinfo{author}{A.~Maisels},
  \bibinfo{author}{T.~Matsoukas}, \bibinfo{author}{C.~Zheng},
\newblock \bibinfo{title}{Analysis of four monte-carlo methods for the solution
  of population balances in dispersed systems},
\newblock \bibinfo{journal}{Powder Technol.} \bibinfo{volume}{173}
  (\bibinfo{year}{2007}) \bibinfo{pages}{38--50}.
\bibitem[{Zucca et~al.(2007)Zucca, Marchisio, Vanni, and Barresi}]{zucca07}
\bibinfo{author}{A.~Zucca}, \bibinfo{author}{D.~L. Marchisio},
  \bibinfo{author}{M.~Vanni}, \bibinfo{author}{A.~A. Barresi},
\newblock \bibinfo{title}{Validation of bivariate {DQMOM} for nanoparticle
  processes simulation},
\newblock \bibinfo{journal}{Ai{ChE} J.} \bibinfo{volume}{53}
  (\bibinfo{year}{2007}) \bibinfo{pages}{918--931}.

\end{thebibliography}

\newpage
\appendix
\section{Moment evolution equations}\label{a:moments}

Equation \eqref{e:rhs} is derived as
\begin{align}
    \int \left( \frac{ \dd P}{ \dd t} = 0 \right) R^l \Rdot^m R_o^n \dd \bx  &\to \label{e:1}\\
	\frac{\partial }{\partial t} \int P R^l \Rdot^m R_o^n \dd \bx + 
	\int \frac{\partial (\Rdot P)}{\partial R} R^l \Rdot^m R_o^n \dd \bx +
 	\int \frac{\partial (\ddot{R} P)}{\partial \Rdot}  R^l \Rdot^m R_o^n \dd \bx &= 0, \label{e:2}\\
	\frac{\partial \mom_{lmn} }{\partial t} + 
	\int \frac{\partial (\Rdot P R^l \Rdot^m R_o^n)}{\partial R} \dd \bx -
	\int \frac{\partial (R^l \Rdot^m R_o^n)}{\partial R} \Rdot P \dd \bx\; &+  \notag\\
 	\int \frac{\partial (\ddot{R} P R^l \Rdot^m R_o^n)}{\partial \Rdot}  \dd \bx -
	\int \frac{\partial (R^l \Rdot^m R_o^n)}{\partial \Rdot}  \ddot{R} P \dd \bx &= 0, \label{e:3}	\\
	\frac{\partial \mom_{lmn} }{\partial t} - l \mom_{l-1,m+1,n} - n \int \Rddot R^l \Rdot^{m-1} R_o^n P \dd \bx &= 0.\label{e:4}
\end{align}
Applying \eqref{e:master} to \eqref{e:1} results in \eqref{e:2}.
Performing integration by parts on the second and third integrals results in \eqref{e:3}.
\eqref{e:4} follows from evaluation of the first and third integrals, application of \eqref{e:mom}, and evaluation of the derivative in the last integrand.
This also matches~\eqref{e:rhs}.

\end{document}